\documentclass[a4paper,12pt,twoside]{article}
\usepackage{a4wide}
\usepackage{graphicx}
\usepackage{dcolumn}
\usepackage{bm}

\def\gtap{\mathrel{ \rlap{\raise 0.511ex \hbox{$>$}}{\lower 0.511ex
   \hbox{$\sim$}}}} 
\def\ltap{\mathrel{ \rlap{\raise 0.511ex
    \hbox{$<$}}{\lower 0.511ex \hbox{$\sim$}}}} 
\newcommand{\bea}{\begin{eqnarray}} 
\newcommand{\eea}{\end{eqnarray}}
\def\beq{\begin{equation}}
\def\enq{\end{equation}}
\def\ba{\begin{eqnarray}}
\def\ea{\end{eqnarray}}
\def\<{<\!\!}
\def\>{\!\!>}
\def\<{\langle}
\def\>{\rangle}

\begin{document}

\input{epsf}

\begin{center}             
\noindent{\Large \tt \bf 
Model-Independent Bound \\[1ex] on the Dark Matter Lifetime
}\vspace{4mm}

\noindent{\large
Sergio Palomares-Ruiz
}\vspace{2mm}

\noindent{\small
IPPP, Department of Physics, Durham University, Durham DH1 3LE,
United Kingdom}

\end{center}

\vspace{6mm}
\renewcommand{\thefootnote}{\arabic{footnote}}
\setcounter{footnote}{0}

\begin{abstract}
If dark matter (DM) is unstable, in order to be present today, its
lifetime needs to be longer than the age of the Universe, $t_U \simeq 4
\times 10^{17}$~s. It is usually assumed that if DM decays it would do
it with some strength through a radiative mode. In this case, very
constraining limits can be obtained from observations of the diffuse
gamma ray background. However, although reasonable, this is a
model-dependent assumption. Here our only assumption is that DM decays
into, at least, one Standard Model (SM) particle. Among these,
neutrinos are the least detectable ones. Hence, if we assume that the
only SM decay daughters are neutrinos, a limit on their flux from DM
decays in the Milky Way sets a conservative, but stringent and
model-independent bound on its lifetime.
\end{abstract}


\section{Introduction} 

It is more than seventy years since an unknown {\sl missing mass} was
first postulated in order to understand the motion of galaxies in
clusters~\cite{Zwicky}. Since then, a lot more evidences at different
scales have been found in favor of the existence of this non-luminous
matter, from the measurements of the rotation curves of galaxies to
the observation of the cosmic microwave background (CMB) (for reviews
see eg. Refs.~\cite{reviews}). Thus the question now is not if the
dark matter (DM) exists, but what is its nature. Many non-baryonic
candidates have been proposed, from the lightest particle in
supersymmetric models, to the lightest Kaluza-Klein excitation in
models of extra-dimensions, to sterile neutrinos, light scalar
particles, axions, particles from little Higgs models, etc.~(see
eg. Refs.~\cite{reviews} for a comprehensive list).

It might well happen that DM consists of different species, with
different interactions with Standard Model (SM) particles and among
themselves. It might also happen that DM was not in thermal
equilibrium in the early Universe, i.\ e.\ it is not a thermal
relic. Nevertheless, in order for it to be present today, there is a 
requirement: it needs to be stable, or at least to have a lifetime
longer than the age of the Universe, $t_U \simeq 4 \times
10^{17}$~s. The possibility of unstable DM is not new and models with
decaying DM have been considered since long ago for different aims in 
astrophysics and cosmology~\cite{decaymodels}. Bounds on its radiative
lifetime have been usually obtained from measurements of the diffuse
gamma ray background~\cite{decayradbounds}. However, although very
stringent, they do not necessarily represent constraints on the actual
DM lifetime, but only set limits on its decay mode into photons. On
the other hand, by evaluating how the expansion rate of the Universe
is affected, bounds on the DM lifetime can be set using CMB
data~\cite{decayboundsCMB}. In this more general case, it is shown that
already the first year of WMAP observations~\cite{WMAP1} constrains DM
lifetime to be longer than $\tau_{\rm{CMB}} = 1.6 \times 10^{18}$~s
at 2$\sigma$ confidence level (CL), with the only assumption that the
decay is into relativistic particles~\cite{decayboundsCMB}. This bound
is rather robust but difficult to improve by further observations, for
the DM decay affects CMB only at large scales, for which errors are
limited by cosmic variance~\cite{decayboundsCMB}. Nevertheless, a
recent study~\cite{decayboundsCMBSN}, which also takes into account
type Ia supernovae (SN) data, improves this limit by about an order of
magnitude, $\tau_{\rm{CMB+SN}} = 2.2 \times 10^{19}$~s at 2$\sigma$
CL.  

In this letter we obtain a lower bound on the DM lifetime, which is
much more restrictive (several orders of magnitude) than that set by
CMB and SN observations and model-independent, unlike that obtained
from observations of the diffuse gamma ray background. 

Among the stable SM particles, neutrinos are the least detectable ones.
Therefore, if we assume that the only SM products from the DM decay
are neutrinos, a limit on their flux, conservatively and in a 
model-independent way, sets a lower bound on the DM lifetime. This is
the most conservative assumption from the detection point of view,
that is, the worst possible case. Any other decay channel (into at
least on SM particle) would produce photons and hence would give rise
to a much more stringent limit. Let us stress that this is not an
assumption about a particular and realistic case. On the other hand,
for the reasons just stated, it is valid for any generic model with
unstable DM, which decays at least into one SM particle. Hence, the
bound so obtained is a bound on the lifetime of the DM particle and
not only on its partial lifetime due to the decay channel into
neutrinos. Thus, following a similar approach to that of
Refs.~\cite{BBM06,YHBA07,PP07}, we consider this case and evaluate the
potential neutrino flux from DM decay in the whole Milky Way, which we
compare with the relevant backgrounds for detection: mainly the well
known and measured atmospheric neutrino flux, which spans over about
seven decades in energy.

\section{Neutrino Fluxes from DM Decay in the Milky Way} 

In what follows we only study DM decays in the Milky Way and do not
consider the diffuse signal from cosmic decays. In general, the latter
is likely to be smaller than, or at most of the same order of, the
former. 

If DM has a lifetime longer than the age of the Universe, $\tau_\chi >
t_U $, the differential neutrino (plus antineutrino) flux per flavor
from DM decay in a cone of half-angle $\psi$ around the galactic
center, covering a field of view $\Delta\Omega = 2 \, \pi \, \left(1 -
\cos\psi \right)$, is given by
\begin{equation}
\frac{ d \Phi}{d E_\nu} = \frac{\Delta\Omega}{4\, \pi} \, {\cal
 J}_{\Delta\Omega} \,\frac{R_{\rm sc} \, \rho_0}{m_\chi \tau_\chi} \,
\frac{1}{3} \, \frac{dN}{dE_\nu} , 
\label{dkflux}
\end{equation}
where $m_\chi$ is the DM mass, $R_{\rm sc}=8.5$~kpc is the solar
radius circle, $\rho_0 = $ 0.3~GeV~cm$^{-3}$ is a normalizing DM
density, which is equal to the commonly quoted DM density at $R_{\rm 
  sc}$, and ${\cal J}_{\Delta\Omega}$ is the average in the field of
view (around the galactic center) of the line of sight integration of
the DM density, which is given by
\begin{equation}
{\cal J}_{\Delta\Omega}  = \frac{2 \, \pi}{\Delta\Omega} \,
 \frac{1}{R_{\rm sc} \, \rho_0}  \, \int_{\cos \psi}^1 \,
 \int_{0}^{l_{\rm max}} \,  \rho (r) \, dl \, d(\cos \psi'), 
\label{Javg}
\end{equation}
where $r = \sqrt{R^2_{\rm sc} -  2 l R_{\rm sc} \cos \psi' + l^2}$,
and the upper limit of integration is 
\begin{equation}
l_{\rm max} = \sqrt{(R_{\rm halo}^2 - \sin^2 \psi R^2_{\rm sc})} +
R_{\rm sc} \cos \psi ~.
\end{equation}
This integral barely depends on the size of the halo $R_{\rm halo}$, as
long as it is larger than few tens of kpc, for the contribution at
large scales is negligible. 

The neutrino (plus antineutrino) spectrum per flavor is given by
$dN/dE_\nu$. If DM is the lightest particle of the new sector, which
is introduced to render a more complete theory, then it can only decay
into SM particles. Hence, the most conservative assumption is that it
decays into neutrino-antineutrino pairs, $\chi \rightarrow \nu \,
\overline{\nu}$. In this case $dN/dE_\nu = 2 \, \delta(E_\nu -
m_\chi/2)$. However, if the lightest particle of the new sector
($\chi_{\rm L}$) is stable, but the next-to-lightest particle
($\chi_{\rm NL}$) is long-lived, the latter could also constitute part
of the DM and decay into the former plus one or more SM particles,
which we conservatively assume to be neutrinos. Commonly, in this
class of models, these two new particles are almost degenerate in
mass, and thus the total energy of the produced neutrinos is
approximately equal to the difference of their masses ($\Delta
M$). For two-body decays, $\chi_{\rm NL} \rightarrow \chi_{\rm L} +
\nu$, $dN/dE_\nu = \delta(E_\nu - \Delta M)$, whereas for three-body
decays, $\chi_{\rm NL} \rightarrow \chi_{\rm L} + \nu +
\overline{\nu}$, the neutrino (and antineutrino) spectrum is
continuous with a maximum energy equal to $\Delta M$. In what follows
we shall consider for concreteness (and for comparison with the CMB
bounds) the case of DM decay into neutrino-antineutrino pairs and
obtain bounds on $\tau_\chi$ as a function of $m_\chi$. From this
calculation, it is straightforward to obtain a bound on the
combination $m_\chi \, \tau_\chi$ as a function of the neutrino
energy, which for the second two-body decay case is equal to $\Delta  
M$. There are two main points to take into account. Whereas in the
first case there is a produced neutrino-antineutrino pair, in the
second there is only one final neutrino (or antineutrino). On the other
hand, in the second case only half of the DM decays (the
next-to-lightest particle of the new sector), for the lightest
particle of the new sector is assumed to be stable, and it also
contributes (usually at comparable level) to the DM. This implies an
overall factor of 4. Finally, although a detailed analysis for
three-body decays is model-dependent, in general this case would give
bounds, for a given neutrino energy, of the same order of magnitude of
those for the two-body decay case. 

In Eq.~(\ref{dkflux}), the factor of 1/3 comes from the assumption
that the decay branching ratio is the same for the three neutrino
flavors. Let us note that this is not a very restrictive assumption,
for even in the case DM decays predominantly into one flavor, there is
a guaranteed flux of neutrinos in all flavors thanks to the averaged
neutrino oscillations between the source and the detector. Hence,
although different initial flavor ratios would give rise to different
flavor ratios at detection, the small differences affect little our
results and for simplicity herein we consider equal decay into all
flavors.

On the other hand, with our definition of ${\cal J}_{\Delta\Omega}$,
all the astrophysical uncertainties in the calculation of the neutrino
flux from DM decays are encoded in ${\cal J}_{\Delta\Omega}$. They
come from our lack of knowledge of the exact DM density $\rho (r)$. As
a matter of fact, the formation of large scale structure is
successfully predicted by detailed N-body simulations which show that 
cold DM clusters hierarchically in halos. The simulated DM profile in
a galaxy like the Milky Way, assuming a spherically symmetric matter
density with isotropic velocity dispersion, can be parametrized as
\begin{equation}
\rho(r) = \rho_{\rm sc} \,
  \left(\frac{R_{\rm sc}}{r}\right)^\gamma \, 
  \left[\frac{1+(R_{\rm sc}/r_{\rm s})^\alpha}{1+
  (r/r_{\rm s})^\alpha}\right]^{(\beta-\gamma)/\alpha},  
\label{rho}
\end{equation}
where $\rho_{\rm sc}$ is the DM density at $R_{\rm sc}$, $r_{\rm s}$
is the scale radius, $\gamma$ is the inner cusp index, $\beta$ is the
slope as $r \rightarrow \infty$ and $\alpha$ determines the exact
shape of the profile around $r_{\rm s}$. 

The three commonly used DM density profiles we
consider~\cite{NFW,Kravtsov,Moore} (see also Ref.~\cite{DMprofiles}) 
tend to agree at large scales, but uncertainties can be significant
in their inner parts. However, for a large field of view, these
uncertainties are much less relevant and do not affect significantly
the calculation of the neutrino flux from DM decay. In addition, and
unlike the case of DM annihilations, the neutrino flux depends on the
line of sight integral of the DM density and not of its square, which
reduces considerably the effect of the inner cusp uncertainty. 

As we will see below, and following a similar approach as in
Ref.~\cite{YHBA07}, we are interested in signals corresponding to
different components of the halo: the full-sky signal and the signal
from a $30^o$ half-angle cone around the galactic center. Whereas for
the former, the value of the average of the line of sight integration
of the DM density, ${\cal J}_{180}$, for the three considered
profiles, can vary at the very most from 1.3 to 8.1, for the latter, 
the value of ${\cal J}_{30}$ might be anything from 3.9 to 24. These
limiting cases are obtained from the range of values for $\rho_{\rm
  sc}$~\cite{BCFS02} which satisfy present constraints from the
allowed range for the local rotational velocity~\cite{v0}, the amount
of flatness of the rotational curve of the Milky Way and the maximal
amount of its non-halo components~\cite{massMW}. For the usually
quoted value of $\rho_{\rm sc}$, for each of the profiles, ($\rho_{\rm
  sc}, {\cal J}_{180}, {\cal J}_{30})=$~(0.27~GeV/cm$^3$, 1.9,
6.5)~\cite{Moore}, (0.30~GeV/cm$^3$, 2.0, 6.1)~\cite{NFW} and
(0.37~GeV/cm$^3$, 2.2, 5.5)~\cite{Kravtsov}. Thus, uncertainties in
the halo profile have fairly small effects on our final results. Here 
we consider the simulation by Navarro, Frenk and White
(NFW)~\cite{NFW} as our canonical profile. From the limiting cases
just discussed, this implies that in the worst scenarios we could be
overestimating (underestimating) the neutrino flux by a factor of
about 1.5 (3.9).

\section{Neutrino Bounds on the DM lifetime} 

For $E_\nu \gtap$~50--60~MeV, the main source of background for a
possible neutrino signal from DM decays is the flux of atmospheric
neutrinos, which has been measured in a number of detectors up to
energies of $\sim$~100~TeV~\cite{atmospheric}. Its spectrum is also
well understood theoretically and different calculations using
different interactions models agree within
20-30\%~\cite{atmostheo,newatmos,FLUKA}. Thus, in order to obtain a
bound on the DM lifetime we need to compare these two fluxes, and in
particular we consider the $\nu_\mu + \overline{\nu}_\mu$ spectra
calculated with FLUKA~\cite{FLUKA}.

Assuming two-body DM decays into neutrino-antineutrino pairs, we first
obtain a general lower bound for $m_\chi \sim$~100~MeV--200~TeV, by
comparing the ($\nu_\mu + \overline{\nu}_\mu$) neutrino flux from DM
decays in the halo with the corresponding atmospheric neutrino flux in
an energy bin of width $\Delta \log_{10} E_\nu = 0.3$ around $E_\nu =
m_\chi/2$. For each value of $m_\chi$, the limit on $\tau_\chi$ is
obtained by setting its value so that the neutrino flux from DM decays
in the Milky Way equals the atmospheric neutrino spectrum integrated
in the chosen energy bin. The reason for choosing this energy bin is
two-fold: on one side, the neutrino signal is sharply peaked around a
neutrino energy equal to half of the DM mass and this choice is within
the experimental limits of neutrino detectors, and on the other side,
for the sake of comparison, we follow the approach of
Ref.~\cite{YHBA07}. Nevertheless, following Ref.~\cite{PP07}, a more
detailed analysis is performed below for $m_\chi \sim$~30--200~MeV.

\begin{figure}[t]
\centerline{\epsfxsize=5.9in \epsfbox{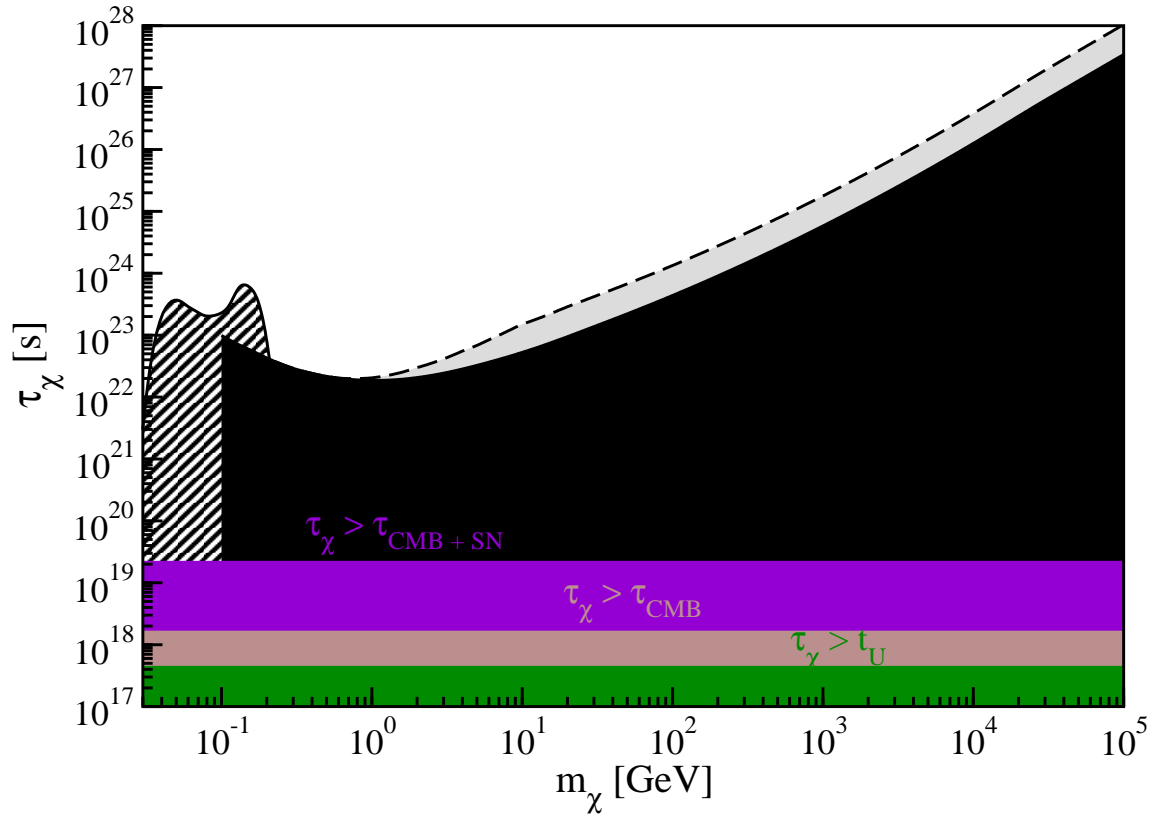}}
\caption{Bounds on the DM lifetime for a wide range of DM masses
  obtained using different approaches: full-sky signal (dark area),
  angular signal (light area) and 90\% CL limit using SK data at low
  energies~\cite{SKSN} (hatched area). Results are obtained for a NFW
  profile and assuming two-body decays into relativistic
  particles (see text). Also shown the bound obtained from CMB
  observations~\cite{decayboundsCMB} and CMB plus SN
  data~\cite{decayboundsCMBSN} (both at 2$\sigma$ CL) and the line
  $\tau_\chi = t_U$.} 
\label{DMbound} 
\end{figure}

The most conservative bound is obtained by using the full-sky signal,
and this is shown in Fig.~\ref{DMbound} where the dark area represents
the excluded region. However, a better limit can be obtained by using
angular information. This is mainly limited by the kinematics of the
interaction. In general, neutrino detectors are only able to detect
the produced lepton and its relative direction with respect to the
incoming neutrino depends on the neutrino energy as $\Delta\theta \sim
30^o \times \sqrt{{\rm GeV}/E_\nu}$. As in Ref.~\cite{YHBA07} and
being conservative, we consider a field of view with a half-angle cone
of $30^o$ ($30^o \times \sqrt{10 \, {\rm GeV}/E_\nu}$) for neutrinos
with energies above (below) 10~GeV. This limit is shown in
Fig.~\ref{DMbound} by the dashed line (light area), which improves
upon the previous case by about a factor of three for $m_\chi
>$~10~GeV.

As anticipated, it is expected that a more detailed analysis, making a
more careful use of the directional as well as energy information for
a given detector, will improve these results. Note for instance that
for energies $\sim$~1-100~GeV neutrino oscillations would give rise to
a zenith-dependent background, roughly speaking a factor of two larger
for downgoing neutrinos as compared to the upgoing flux, whereas we
expect a nearly flat background for other energies for which
oscillations do not take place. On the other hand, the signal from DM 
decays in the halo is expected to change by a factor of $\sim$~7 for a
half-angle cone of 30$^o$ around the galactic center as compared to
the signal within the same field of view but coming from the opposite
direction. Thus, making use of the directional information would
certainly render more stringent bounds. Nevertheless, and although a
detailed and detector-dependent analysis is beyond the scope of this
letter, we show how such a more careful treatment of the energy
resolution and backgrounds can substantially improve these limits. For
this and also extending the bounds to lower DM masses, we consider the
low energy window below $\sim$~100~MeV (i.\ e. for $m_\chi \ltap
200$~MeV) and perform an analogous analysis to that in
Ref.~\cite{PP07}.
 
In this energy range the best data comes from the search for the
diffuse supernova background by the Super-Kamiokande (SK) detector
which has looked at positrons (via the inverse beta-decay reaction,
$\overline{\nu}_e + p \rightarrow e^+ + n$) in the energy interval
18~MeV--82~MeV~\cite{SKSN}. As for these energies there is no
direction information, we consider the $\overline{\nu}_e$ signal  
coming from the whole sky. In this search, the two main sources of
background are the atmospheric $\nu_e$ and $\overline{\nu}_e$ flux
and the Michel electrons and positrons from the decays of
sub-threshold muons. Below 18~MeV, muon-induced spallation products
are the dominant background, and below $\sim$~10~MeV, the signal
would be buried below the reactor antineutrino background.

Although for $E_\nu \ltap$~80~MeV the dominant interaction is the
inverse beta-decay reaction (with free protons), the interactions of
neutrinos (and antineutrinos) with the oxygen nuclei contribute
significantly and must be considered. For our analysis we have
included both the interactions of $\overline{\nu}_e$ with free protons
and the interactions of $\nu_e$ and $\overline{\nu}_e$ with bound
nucleons, by considering, in the latter case, a relativistic Fermi gas
model~\cite{SM72} with a Fermi surface momentum of 225~MeV and a
binding energy of 27~MeV. We then compare the shape of the background
spectrum to that of the signal and perform a $\chi^2$ analysis so that
we can extract the limit on the DM lifetime in an analogous way as it
was done to obtain an upper bound on the annihilation cross section
for the case of DM annihilation in Ref.~\cite{PP07}, where we refer
the reader for all the details of this analysis~\footnote{Note that
  there is an error in Eq.(8) of Ref.~\cite{PP07}, which should read
  $P(\alpha) = K \cdot e^{-\chi^2/2}$. Nonetheless, this implies very
  small corrections to the results presented. I thank O.~L.~G.~Peres
  for pointing this out.}. The 90\% CL limit is shown in
Fig.~\ref{DMbound} by the hatched area and it clearly improves (and
extends to lower masses) by up to an order of magnitude upon the
general and very conservative bound obtained with the simple analysis
described above.

Finally, let us note that in principle, if the DM mass is not known, 
DM annihilation and DM decay in the halo might have the same
signatures. However, whereas the decay signal depends linearly on the 
DM halo density, the annihilation signal depends on its square. Hence,
in case of a positive signal, directional information is crucial to
distinguish between these two possibilities.

\section{Conclusions} 

In this letter we have obtained a general bound on the DM lifetime,
which is several orders of magnitude more stringent than previous
limits~\cite{decayboundsCMB,decayboundsCMBSN}. In order to do so, we
have considered that the only SM products from DM decays are neutrinos,
which are the least detectable particles of the SM. Thus, regardless
of how likely this is, by making this assumption we can obtain a
conservative but model-independent bound on the DM lifetime. To do so
we have considered the flux of neutrinos coming from DM decays in the
Milky Way for an energy interval from $\sim$~50~MeV to $\sim$~100~TeV
and have compared it to the dominant background, the well known and
measured atmospheric neutrino flux. For concreteness we only show
results for two-body DM decays into relativistic SM particles,
although it is straightforward to generalize this result to other
two-body decays. On the other hand, the model-dependent case of
three-body decays should render limits of the same order of
magnitude. We have obtained a general bound by considering the signal
from the whole Milky Way and imposing that it has to be at most equal
to the background in a given energy interval. We have also shown how
this crude, but already very stringent limit, can be substantially
improved by more detailed analysis which make careful use of the
angular and energy resolution of the detectors, as well as of
backgrounds. In this way, following the analysis of Ref.~\cite{PP07},
we have obtained the 90\% CL lower bound on the DM lifetime for
$m_\chi \sim$~30--200~MeV, which is an order of magnitude more
stringent.

In summary, we have shown that neutrinos can be used to set very
stringent and model-independent bounds on the DM lifetime, with the
only assumption that if DM is unstable, it decays into at least one SM
particle. As our main result, we have improved by several orders of
magnitude upon previous limits~\cite{decayboundsCMB,decayboundsCMBSN}.

\vspace{-2mm}
\section*{Acknowledgments}
The author thanks G.~Battistoni for providing him with the atmospheric
neutrino fluxes, J.~Beacom and T.~Montaruli for helpful comments and
Y.~Santoso for discussions. SPR is partially supported by the Spanish
Grant FPA2005-01678 of the MCT.

\end{document}